\documentclass[12pt,preprint]{aastex}

\newcommand{\bdv}[1]{\mbox{\boldmath$#1$}}

\def\au{{\rm au}} 
\def\sinc{{\rm sinc}} 
\def\kms{{\rm km}\,{\rm s}^{-1}}

\def\kpc{{\rm kpc}}
\def\mas{{\rm mas}}

\def\min{{\rm min}}

\def\e{{\rm E}}

\def\bmu{{\bdv\mu}}

\def\btheta{{\bdv\theta}}

\begin{document}
\title{One Small Step for {\it Roman}; One Giant Leap for Black Holes}

\author{\textsc{
Andrew Gould$^{1,2}$
}}

\affil{$^{1}$Department of Astronomy, Ohio State University, 140 W.
18th Ave., Columbus, OH 43210, USA}

\affil{$^{2}$Max-Planck-Institute for Astronomy, K\"{o}nigstuhl 17,
69117 Heidelberg, Germany}

\begin{abstract}

  The {\it Roman} microlensing program can detect and fully
  characterize black holes (BHs) that are in orbit with about 30
  million solar-type and evolved stars with periods up to the mission
  lifetime $P\la T\sim 5\,$yr, and semi-major axes $a\ga 0.2\,\au$,
  i.e., $P\ga 10\,$d $(M/M_\odot)^{-1/2}$, where $M$ is the BH mass.
  For BH companions of about 150 million later (fainter) main-sequence
  stars, the threshold of detection is $a\ga 0.2\,\au \times
  10^{(H_{\rm Vega}-18.5)/5}$.  The present {\it Roman} scheduling
  creates a ``blind spot'' near periods of $P\sim 3.5\,$yr due to a
  2.3 year gap in the data.  It also compromises the characterization
  of BHs in eccentric orbits with periods $P\ga 3\,$yr and peribothra
  within a year of the mission midpoint.  I show that one can greatly
  ameliorate these issues by making a small adjustment to the {\it
    Roman} observing schedule.  The present schedule aims to optimize
  proper motion measurements, but the adjustment proposed here would
  degrade these by only 4--9\%.  For many cases of $P\ga 90\,$d BHs,
  there will be discrete and/or continuous degeneracies.  For G-dwarf
  and evolved sources, it will be straightforward to resolve these by
  radial-velocity (RV) follow-up observations, but such observations
  will be more taxing for fainter sources.  Many BH-binaries in orbits of
  $5\,{\rm yr} \la P \la 10\,{\rm yr}$ will be reliably identified
  as such from the {\it Roman} data, but will lack precise orbits.
  Nevertheless, the full orbital solutions can be recovered by combining
  {\it Roman} astrometry with RV followup observations.  BH binaries with
  periods $10\,{\rm yr}\la P\la 95\,{\rm yr} (M/10 M_\odot)^{1/4}$ can be
  detected from their astrometric acceleration, but massive multi-fiber
  RV monitoring would be needed to distinguish them from the
  astrophysical background due to stellar binaries.  
  
\end{abstract}


\section{{Introduction}
  \label{sec:intro}}

To date, a total of four quiescent stellar-mass black holes (BHs) are known,
one that is actually isolated \citep{ob110462a,ob110462b,ob110462c},
and three found from {\it Gaia} observations of stars that orbit the BHs
in relatively wide ($1.5\,\au \la a \la 17\,\au$)
orbits \citep{elbadry23a,elbadry23b,mazeh24}.


This handful of detections already clearly suggests that BH characteristics
are a strong function of environment.  For example, three of the four BHs have
masses of ${\cal O}(10\,M_\odot)$, local kinematics, and (for the two with
measured abundances) slightly subsolar metallicities, while the fourth
has $M=33\,M_\odot$, halo kinematics, and [Fe/H]$=-2.56$.

Hence, by probing BHs toward and in the Galactic bulge, one would likely
gain new insights into these objects.

High-cadence, high-precision long-duration (i.e., microlensing-style)
observations of the Galactic bulge permit two channels to discover and
characterize BHs.  The first is to detect these BHs as lenses
\citep{gouldyee14}.  The microlensing light curve then directly yields
the so-called ``microlens parallax'' $\pi_\e$.  To extract the BH
mass, $M=\theta_\e/\kappa \pi_\e$ (where $\kappa =
8.14\,\mas/M_\odot$), one must also measure the Einstein radius,
$\theta_\e$.  \citet{gouldyee14} proposed that $\theta_\e$ could be
measured from astrometric microlensing \citep{walker95,hnp95,my95}
from the survey itself, and this was further explored by
\citet{kb222397}.  However, \citet{industrial} argued that while this
technique could work well for BHs in the disk (along the line of sight
to bulge sources), it would be very challenging for bulge BHs because
both $\pi_\e$ and $\theta_\e$ tend to be small.  Regarding $\pi_\e$,
\citet{industrial} argued that solar-orbiting-satellite microlens
parallaxes (rather than annual parallax) would be needed to robustly
measure $\pi_\e$ for most bulge BHs.  Regarding $\theta_\e$, he argued
that interferometric resolution of the lens images
\citep{delplancke01,kojima1,cassan22} would be needed.

Regardless of whether bulge BHs can be detected as lenses, the same
surveys open the possibility of a second channel: BHs detected from
the astrometric motions of stars that orbit them.  Assuming that the
photon-limited statistics that were analyzed by \citet{gould15} for
{\it Roman} can be achieved, then it will be straightforward to detect
and characterize such BH-star systems with periods $10\,{\rm d} \la P
\la T=5\,$yr, down to $H_{\rm Vega}\sim 18.5$ (i.e., G dwarfs).
Here $T=5\,{\rm yr}$ is the nominal lifetime of the mission.
More precisely, the
short period limit is set by a minimum semi-major axis $a\ga 0.2\,\au$,
(so, $P\sim 10\,{\rm d} (M/10 M_\odot)^{-1/2}$) at
which point, the source orbital motion will have an amplitude
that is only five times smaller than its parallax, which latter will be
measured with $50\,\sigma$ precision.  The long-period limit is set by
the need to confidently measure $P$ and $a$,
from the astrometric data from the $T=5\,$yr mission.  If $P$ and $a$
are known, then the mass function $M_* = M_\odot[(a/\au)^3/(P/{\rm
    yr})^2]$, and any value of $M_*$ that is more than a few $M_\odot$
will clearly be a BH.  BHs can also be detected in orbit with fainter
stars but with a lower limit $a\ga 0.2\,\au\times 10^{(H_{\rm Vega}-18.5)/5}$
due to reduced signal-to-noise ratio (SNR).

Unfortunately, the present {\it Roman} observing strategy is comprised
of two reasonably well covered intervals, each lasting $\sim 1.2$ years,
that are separated by a gap of $\sim 2.3$ years.  This artificially-induced
gap compromises the detection of a relatively broad range of periods
centered on $P=3.5\,$yr.  That is, at exactly $P=3.5\,$yr, the two
$1.2\,$yr intervals cover exactly the same phase of the orbit, greatly
reducing the strength of the signal (so, chance that the system will
be detected) and rendering orbit reconstruction much more difficult.

According to reliable rumor mills, the densely covered {\it Roman} microlensing
seasons (each $\Delta t =72\,$d, with about 90 observations per day),
may be supplemented with the 4 remaining ``off-seasons'' being thinly
covered (once per day).  Relative to the relevant periods $P\sim 3.5\,$yr,
these off-seasons are short, $\Delta t/P\sim 6\%$ and so for present
purposes can be considered as a single point, with error bar
$\sigma=\sigma_0/\sqrt{\Delta t/{\rm d}}\sim 60\,\mu$as, where
$\sigma_0\sim 500\,\mu$as is the precision at $H_{\rm Vega}=18.5$.
This compares to an angular semi-major axis
$a/D_s\simeq a/R_0 = 620\,\mu{\rm as}(M/10 M_\odot)^{1/3}(P/3.5\,{\rm yr})^{2/3}$.
Four such 10-$\sigma$ measurements over the $2.3\,$yr ``gap'' in the main
survey would almost certainly resolve the gap-induced degeneracies.  However,
as this adjustment has not yet been adopted, I will ignore these
extra observations in what follows.  The utility of my proposed revision
should ultimately be assessed in the context of the full observing strategy.
Note, in particular, that the advantages of my revised observing schedule
for detecting BH-binaries in very wide orbits, $P\la 100\,{\rm yrs}$, are
not affectected by the incorporation (or lack thereof) of additional low-cadence
observations.  See Sections~\ref{sec:expected-signal} and \ref{sec:background}.

\section{{Proposed Modification to the Observing Schedule}
  \label{sec:modify}}

I model the {\it Roman} observations as taking place during $m=6$ campaigns,
each centered on an equinox, each consisting of $n\gg 1$ observations that
are uniformly distributed over a total time of $\Delta t=72\,$d.  That is,
the observations are at times $t_{i,j}$ 
\begin{equation}
  t_{i,j} = \tau_i + \delta\tau_j, \qquad \delta\tau_j\equiv
  {j\over n} \Delta t;\qquad (-n/2 < j < n/2).
\label{eqn:tij}
\end{equation}
Note that there are a total of $N=nm$ observations.
I advocate the following change for the six $\tau_i$
\begin{equation}
\left(\matrix{\tau_i^{Roman}\cr \tau_i^{\rm Gould}}\right) =
  \left(\matrix{-9 & -7 & -5 & +5 & +7 & +9\cr
    -9 & -7 & -1 & +1 & +7 & +9}\right)Q;\qquad
Q\equiv{{\rm yr}\over 4}.
  \label{eqn:roman-vs-ag}
\end{equation}

\begin{figure}
\plotone{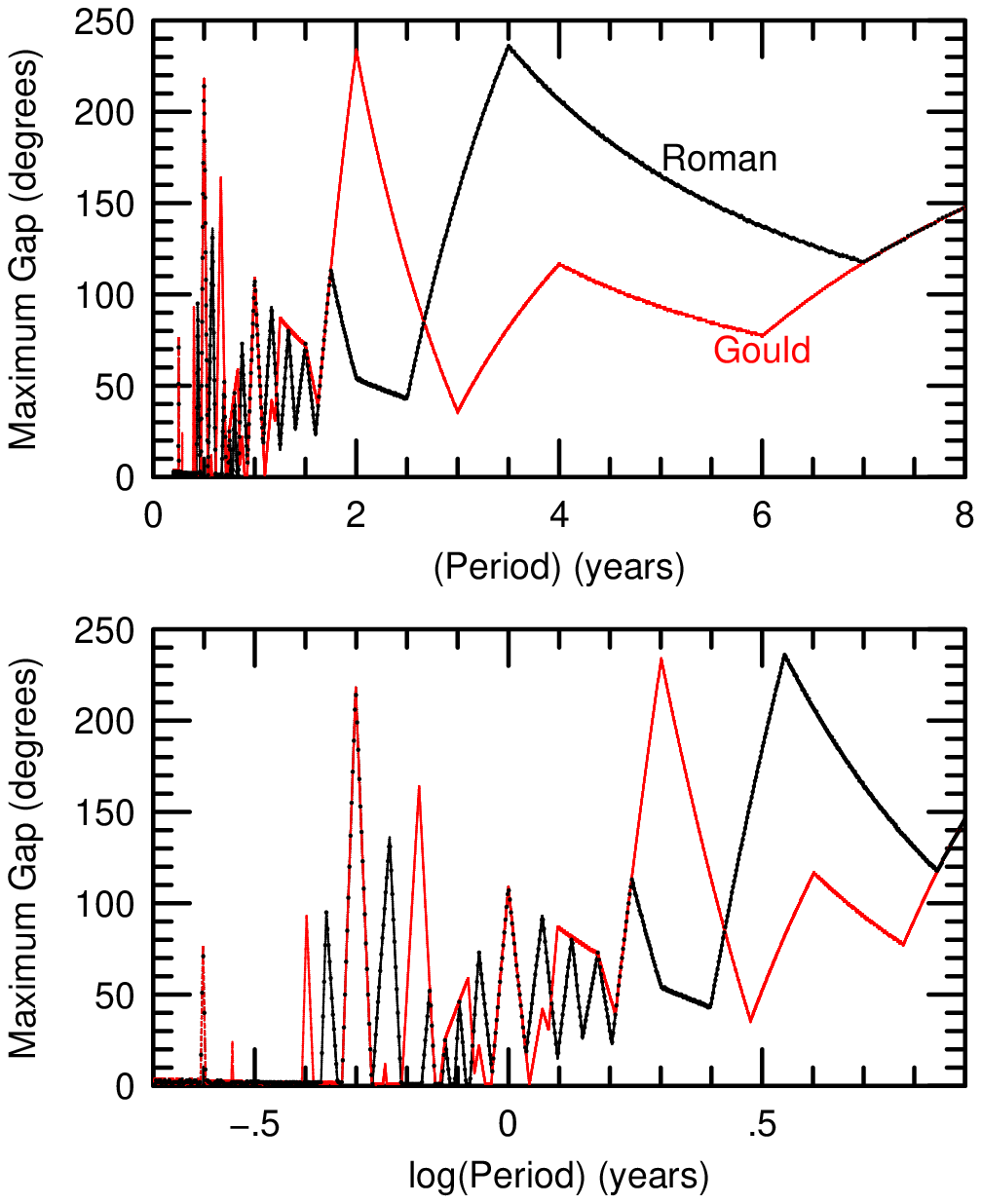}
\caption{Largest gaps of folded data as a function of period, $P$, for
  {\it Roman} data under the current observing schedule (black,
  ``Roman'') and the schedule proposed in the current work (red,
  ``Gould'').  See Equations~(\ref{eqn:tij}) and
  (\ref{eqn:roman-vs-ag}). Gaps are hardly an issue for periods $10\,{\rm
    d}\la P \la 145\,{\rm d}$, and they play only a modest role for
  $145\,{\rm d}\la P \la 600\,{\rm d}$.  However, at longer periods,
  they can be a major issue, particularly for the ``Roman'' schedule
  for periods $P\ga 3\,$yr.  Note that the ``Roman'' schedule is
  replotted with slightly larger points and much more sparsely, so
  that the regions where the two curves are identical can be easily
  identified.  }
\label{fig:gaps}
\end{figure}

The main reason for the change is to remove artificial gaps in the
sensitivity to BHs, but it does have other implications, which must
be addressed.  Figure~\ref{fig:gaps} shows the largest phase-gap as a
function of period under the two strategies.  The current {\it Roman}
strategy is plotted in black and my (Gould) strategy is plotted ``on top''
of this in red.  Because the two curves perfectly overlap in some regions,
I replot the {\it Roman} strategy on the very top, but with much sparser
sampling and slightly larger points.

Figure~\ref{fig:gaps} shows that there are five features that the two
strategies have strictly in common.  First, there are no gaps for
$P<72\,$days because each strategy is composed of six 72-day
continuous seasons.  Second, there is a nearly triangular peak of
$218^\circ$ at $P=0.5\,$yr and with a base of 25 days.  At the peak
period, each season covers exactly the same range of phases, and thus
always contains the intra-equinox gap of 110.5 days, corresponding to
$218^\circ$.  Both strategies cover the first and last available
equinoxes, so in either case, as $P$ deviates from exactly 0.5 yr, the
last season moves relative to the first to progressively cover the
gap.  Third, there is another nearly triangular peak of $109^\circ$ at
$P=1\,$yr (i.e., exactly twice the previous one) and with a base of 64
days.  The reason behind this is almost identical, but applied to the
fact that both strategies observe the first two and last two
equinoxes.  The largest gap is still the one between two successive
seasons, but this corresponds to half the angle.  Fourth, there is yet
another peak at $P=0.25\,$yr, but this has a base of only 2 days and so
has no practical significance.  Finally, at long
periods, $P>7$ years, the two strategies have the same rising gap
size.  This is because they both cover the first and last equinoxes,
so the largest gap is simply ($P-4.7\,$yr) in both cases.

The two strategies also have two features that are qualitatively very similar.
First, each contains two triangular features with peaks of ${\cal O}(120^\circ)$
between $P=0.4\,$yr and $P=0.67\,$yr (in addition to the one at $P=0.5\,$yr).
Second, they have a series of structures of height ${\cal O}(75^\circ)$
between $P=0.7\,$yr and $P=1.6\,$yr (in addition to the one at $P=1\,$yr).
For the {\it Roman} strategy, these are a set of regular peaks, while for the
Gould strategy, they are irregular.  In sum, the two strategies have
qualitatively similar sensitivity for $P<1.6\,$yr.

Overall, the main difference between the two is that the {\it Roman}
strategy has a peak at $P=3.5\,$yr whereas the Gould strategy has a
peak at $P=2.0\,$yr.  In both cases, the peak is at $235^\circ$ and in
both cases the peak period is equal to the periodicity of the
observations.  However, the Gould peak is narrower in $\log P$ (and
even more so in linear $P$), and that is why I consider it to be
better.

\section{{BHs in $P\la T=5\,$yr Orbits}
  \label{sec:5yr}}

The central point of the current work is that, under the assumption that
root-$N$ statistics apply to the {\it Roman} astrometric measurements
(see discussion in \citealt{gould15}),
the survey will automatically probe of order 100 million stars for BH binaries
in orbits $P\la T=5\,$yr.  This is obvious, once it is pointed out.
Nevertheless, there are a few points and sanity checks that must be made.

\subsection{{Minimum Information Content}
  \label{sec:mini-info}}

The first point is that from the point of view of information flow, there
are, effectively, only $m=6$ independent epochs for the case of
$2\,{\rm yr}\la P \la 5\,{\rm yr}$.  These result in exactly 12
independent measurements, i.e., two astrometric coordinate measurements
for each epoch.  This exactly equals the total number of parameters to be
fit, namely the five position-parallax-proper-motion
(pppm) parameters $(\btheta_0,\pi,\bmu)$ and the seven
Kepler parameters\footnote{Another way to count parameters is to note
that 14 parameters are required to completely specify the system, i.e.,
one mass and six phase-space coordinates at a given instant, $t_0$,
for each of the two bodies.  The 12 phase-space coordinates can be expressed
as 6 system coordinates and 6 relative coordinates.  Among these 14 parameters,
there is no information in the astrometric data about the mass of the source
(which is estimated photometrically) nor the system radial velocity (which
remains completely indeterminate).  This leaves 12 parameters to be fit.
}.  Hence, there is enough information to support the
fit, but no redundancy.

There are two reasons that redundancy is usually considered to be desirable.
First, even if the number of measurements is equal to the number of parameters,
there can still be discrete degeneracies and/or highly correlated parameters,
and additional observations can break these degeneracies and/or constrain
the correlations.  In the present case, I will show that resolution of
these degeneracies will require radial-velocity (RV) followup observations
(assuming that the survey itself is not extended beyond its current $T=5\,$yr
mandate).

The second reason that 
redundancy is usually desirable is to guard against catastrophic errors
in one or more of the measurements.  However, in this case, each of the $m=6$
measurements is, in effect, an average of $n> 6000$ individual epochs.  Hence,
catastrophic measurement errors play no role.

Now, if there were, for example, only $m=5$ measurements (say, because of
satellite failure after 5 microlensing seasons) then there would still
formally be a solution because there would in fact be $N>30000$ points,
not just 6.  However, because $\Delta t\ll T$ and even $\Delta t\ll T/m$,
the fact that the $n$ points are not literally all at the same epoch but
rather are spread out over $\Delta t$, means that the 12 parameters of the
fit would be highly correlated.  Nevertheless, even if such a catastrophe does
come to pass, it will in many cases still be possible recognize strong BH
candidates with periods $P\la T$, and these can be followed up with RV
observations.
The typical internal velocities of these systems will be
$v = 38\,\kms (M/10\,M_\odot)^{1/3} (P/5\,{\rm yr})^{-1/3}$.
Even after allowing for the $\sin i$ reduction of this amplitude, RV
measurements of good precision should be feasible on large telescopes.

\subsection{{$2\,{\rm yr}\la P \la 5\,{\rm yr}$}
  \label{sec:2-5}}

Second, the detection process will be very different in different period
regimes, at least at a conceptual level.  As just mentioned, in the
$P\ga 2\,$yr regime, it will be possible to bin the data from each season
with very little loss of information, and then rapidly fit the 12 data points
to 12 parameter-models.  If we restrict consideration to complete orbits
$P\la T$, then the overwhelming majority of such fits will reveal ordinary
stellar binaries, or binaries composed of a star together with either
a neutron star (NS) or a white dwarf (WD).  However, no matter how
numerous these ``contaminants'' are, they cannot be confused with BHs
(provided that the fit returns approximately the true parameters of the
system)
because each solution will yield a definite mass function $M_*$, which
combined with the estimate of the mass of the source, $M_s$, will yield
a mass estimate of the companion:
\begin{equation}
  M=q M_s; \qquad {q^3\over (1+q)^2}\equiv {M_*\over M_s}.
  \label{eqn:massfunction}
\end{equation}
For $M_*\gg M_s$, modest errors in the estimate of $M_s$ play only a small role
because $M=qM_s\rightarrow M_* + 2 M_s$.

It is important to point out that for dark companions (including BHs and NSs),
the value of $M_*$ inferred from the parameters $a$ and $P$ from the
astrometric solution of the orbit, will be the true $M_*$.  By contrast,
for luminous companions, the inferred $a$ will be smaller than the true $a$
leading to an underestimate of $M_*$.  However, this can never lead to
a stellar system being confused with a BH-system because the companion mass
will be underestimated, not overestimated.

On the other hand if the source is blended with either a random field
star or a companion in a much wider orbit, then $a$ (and so $M_*$) will
be underestimated, leading to an underestimate of $M$ via
Equation~(\ref{eqn:massfunction}).  This can in principle
lead to the BH being missed altogether.  However, if the BH is included
in the sample, then its mass can be checked (and corrected) by making
a few RV measurements.  If $a$ has been underestimated due to unrecognized
blending, then the amplitude
of RV variations will be underestimated by exactly the same factor.

\subsection{{$10\,{\rm d}\la P \la 0.5\,{\rm yr}$}
  \label{sec:10d-0.5yr}}

At the shortest periods $P\ll 1\,$yr, the 5 pppm parameters can be
determined from a standard fit to the 6 binned points, and then the
resulting pppm model can be subtracted from each of the $N$ data
points.  Because of the high cadence, $\Gamma=4\,{\rm hr}^{-1}$,
the number of points per period is also very high,
$\Gamma P\sim 900 (P/10\,{\rm d})$.  Hence, they can be binned in
time intervals $\delta t=P/\eta$, where $\eta\sim 50$ would be
a conservative choice.

The resulting astrometric series can be folded in period
steps $\Delta P$, such that $\Delta P/P =  \delta t/T$, i.e.,
$\Delta P = P\delta t/T = P^2/\eta T$, and then fit for a
7-parameter Kepler solution.  This implies an effective number of trials
for each star of $\int_{P_\min} dP/\Delta P = \eta T/P_\min =10^4$, where
$P_\min=10\,$d.  Thus, the total number of trials for all stars in the field
is $N_{\rm trials} \sim 10^{12}$, which implies a threshold for detection
of at least $\Delta\chi^2_\min = 2\ln N_{\rm trials} = 55$.  Formally, a
10-$\sigma$ detection criterion satisfies this condition, but actual
practice may show that a higher threshold is needed in the short-period
regime.  

In this regime, the problem of ``contamination'' by ``ordinary'' binaries
(i.e., with stellar, WD, or NS companions)
will be orders of magnitude less severe
compared to Section~\ref{sec:2-5}.
The very shortest periods that are accessible to BH binaries
will be completely free of such ``contaminants''.
That is, I have set the lower limit
at $a_{\min,0}=0.2\,\au$ for $M=10\,M_\odot$ (so $P=10\,$d) for
$H_{\rm Vega}=18.5$
sources in order to ensure a 10-$\sigma$ detection.  For a NS of mass
$M=1.4\,M_\odot$ and a $1\,M_\odot$ source, the same requirement yields
$a_{\min,\rm NS} = (2.4/1.4)\,a_{\min,0} = 0.34\,\au$ and so a period
$P_\min = 47\,$d.  In fact, NS companions will be relatively rare, and
the most common dark companions will be WDs ($\sim 0.7\,M_\odot$).
In addition there will be stars of similar masses to WDs that, while not
truly ``dark'', are sufficiently less luminous than the source that
they do not significantly debase the astrometric signal.  A similar
calculation yields
$a_{\min,\rm WD} = 0.49\,\au$  and $P_\min = 95\,$day.
This limiting period is already outside the short-period regime that
I am discussing here.

Therefore, the short-period regime will be almost entirely free of
``contaminants'', except for some NS, which are themselves very interesting
objects.  Note also that detections in this regime are the most easily
confirmed by RV followup:  all that is needed is two epochs separated
by half a period, whose RV difference can be predicted from the astrometric
model, and will be many tens of $\kms$, except in pathological cases.
Hence, the search can probably be pushed very close to the statistical limits
in this regime.

\subsection{{$0.5\,{\rm yr}\la P \la 2\,{\rm yr}$}
  \label{sec:0.7-2}}

For the intermediate regime, $0.5\,{\rm yr}\la P \la 2\,{\rm yr}$,
an intermediate approach will be needed, e.g., perhaps binning the
data in 12-day bins, and then fitting the resulting 36 points to
12-parameter models.

\subsection{{Discrete and Continuous Degeneracies}
  \label{sec:degen}}

The form of the observational schedule (whether Gould or {\it Roman})
leads to two types of degeneracies.  First, because there are gaps in
the orbital coverage of the folded data, there can be a continuous
set of orbital parameters that yield essentially equally good fits to the
data.  The most extreme example in the current context would be the
$235^\circ$ coverage gaps shown in Figure~\ref{fig:gaps} at the
$P=2.0\,$yr and $P=3.5\,$yr peaks for Gould and {\it Roman} respectively.
With only 35\% of the orbit covered, it is likely that little more can
be extracted than the vector acceleration and jerk, i.e., 4 parameters.
Hence, 7-Kepler-parameter fits would be highly unconstrained.

\begin{figure}
\plotone{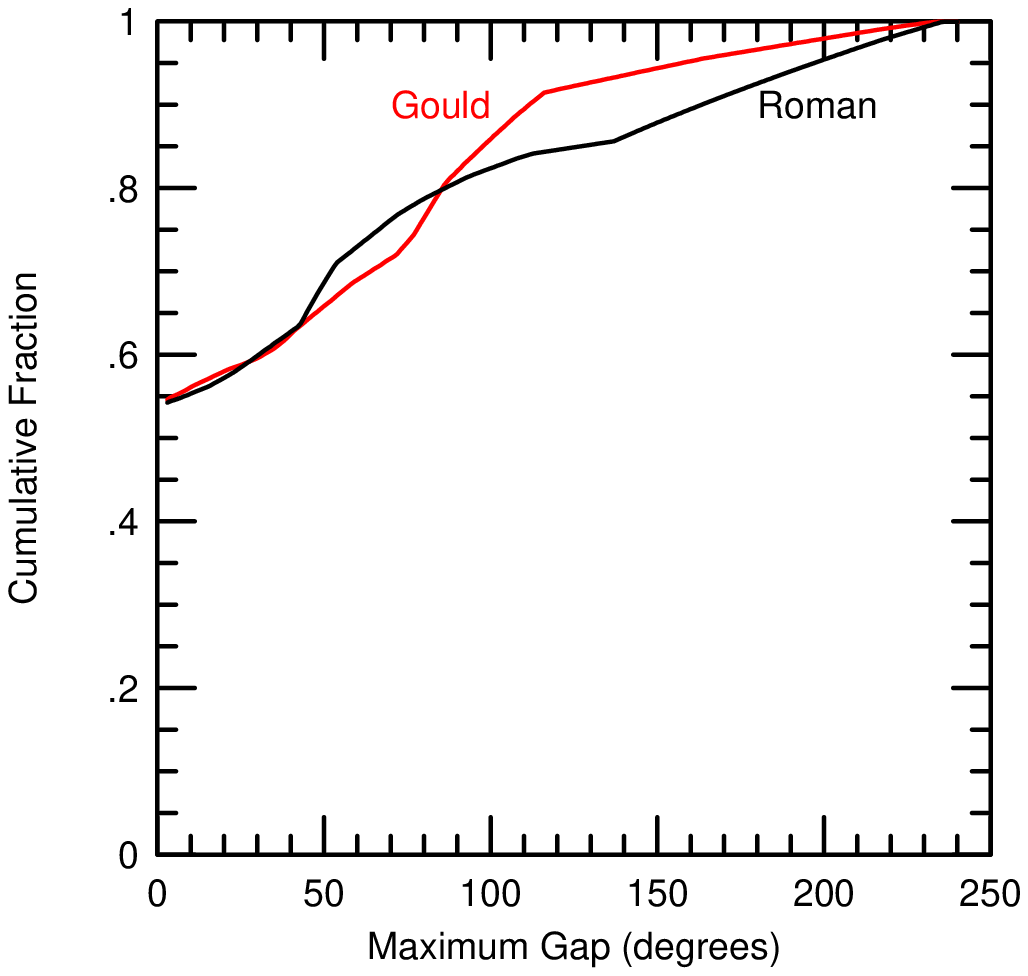}
\caption{Cumulative distribution of the largest gaps (in degrees) for the
  current (black, ``Roman'') and proposed (red, ``Gould'') observing
  schedules, as given by Equations~(\ref{eqn:tij}) and
  (\ref{eqn:roman-vs-ag}).  A majority (by log period, $P$), i.e., 55\%,
  have no gaps, and 80\% have gaps $\la 90^\circ$.  However, large
  gaps are a major issue at long periods, particularly for the
  ``Roman'' schedule.  See Figure~\ref{fig:gaps}.}
\label{fig:cumul}
\end{figure}

However, Figure~\ref{fig:cumul}, which shows the cumulative fraction
of log periods ($10\,{\rm d}< P < 6\,{\rm yr}$) as a function of
largest gap, demonstrates that such extremely poor orbit coverage is
rare.  First, 55\% of log periods have no gaps, while for 80\%, the
largest gap is $<90^\circ$.  Even at this boundary, while there could
be significant uncertainty in the eccentricity if the peribothron fell
near the center of the gap, it is unlikely that the range of allowed
mass functions, $M_*$, would extend beyond the BH regime.  As the gap
increases, problems will accumulate, but these affect only a small
fraction of all BH within the 2.35 decades of periods being probed.
Those systems that are classified as BH binaries with reasonable
confidence, can be followed up with RV, which will resolve all degeneracies.
The small fraction that cannot will be lost, but this is just ``the normal
cost of doing business'' in astronomy.
(Note that 36.7\% of this log-space lies in the range
$10\,{\rm d}<P<72\,{\rm d}$
where there are no gaps because the observing seasons last 72 days, while
another 12.9\% lie in the range $72\,{\rm d}<P<144\,{\rm d}$, where there
are hardly any gaps.)

The second type of degeneracy is aliasing due to the periodic structure of
the observations.  If the observation schedule has gaps at period intervals
$P_{\rm samp}$ (e.g., the sidereal day or the year in ground-based observations),
then when the data are folded by trial periods, aliased periods $P_k$ can
fit the data nearly as well as the true period $P_{\rm true}$ provided
that
\begin{equation}
  {1\over P_k} - {1\over P_{\rm true}} = {k\over P_{\rm samp}},
  \label{eqn:alias}
\end{equation}
where $k$ can be either positive of negative, but usually $|k|$ is small.

Before analyzing to what extent this phenomenon can be at play in
the {\it Roman} data set, I first comment on what the impact would be
if and when it occurred.  The nature of this degeneracy is that the final
orbit ``looks'' the same, so that, in particular, the semi-major axis
would be the same.  Hence, the mass function $M_*$ would change 
as $P^{-2}$.  Given that all BH detections will be checked by RV,
the only real concerns are that the aliasing would drive $M_*$ out of
the BH regime, or that it would drive large numbers of stellar binaries
into the BH regime, thereby overwhelming RV resources.

Specifically, we have
\begin{equation}
  {M_{*,k}\over M_{*,\rm true}} = \biggl({P_k\over P_{\rm true}}\biggr)^{-2}=
  \biggl(1 + k{P_{\rm true}\over P_{\rm samp}}\biggr)^{2}.
  \label{eqn:alias-mstar}
\end{equation}
Thus, if $P_{\rm samp}\gg P_{\rm true}$, then there are no major concerns, but
if $P_{\rm samp}/P_{\rm true}\la$ a few, then the issue must be considered
seriously.

The Gould schedule has three periodicities.  First, the core observational
sequence, [0.2 yr ON, 0.3 yr OFF, 0.2 yr ON, 1.3 yr OFF], is repeated
every $P_{\rm samp}=2\,$yr, for a total of three times.  Second, the pair of
annual ($P_{\rm samp}=1\,$yr)
equinoxes [0.2 yr ON, 0.3 yr OFF, 0.2 yr ON, 0.3 yr OFF], can be
regarded as repeating each year (with no observations in alternate years).
Finally, the equinoxes themselves can be regarded as a ($P_{\rm samp}=0.5\,$yr)
cycle that is just not observed during 4 of the 10 cycles.  However,
as the frequencies (inverse periods) of these three are all multiples of
the first, we can simply consider Equation~(\ref{eqn:alias}) with
$P_{\rm samp}=2\,$yr.

There is no need to be concerned about aliasing for $P_{\rm true}<\Delta t=72\,$d.
The main reason is that $P_{\rm true}/P_{\rm samp}<\Delta t/P_{\rm samp} =0.1$
is small, so according Equation~(\ref{eqn:alias-mstar}), $M_*$ cannot
be aliased into or out of the BH range.  In addition, toward the upper
end of this range, each season will have large enough SNR to yield a solution
based on that season alone, during which there is continuous sampling, while
toward the lower limit $P_{\rm true}/P_{\rm samp}$ is yet several times smaller
than the above limiting calculation.

For periods $P\ga T/2=2.5\,$yr, there also cannot be any issue with
aliasing because the astrometric series is not completely folded even once.

However, I believe that aliasing may become an issue for intermediate
periods, $0.5\,{\rm yr}\la P\la 2.5\,$yr.
This would have to be investigated by simulations, which are beyond the
scope of the present work.

\subsection{{Remarks on Luminous (Giant and Sub-giant) Sources}
  \label{sec:giants}}
  
I should also mention the special case of luminous sources, i.e.,
subgiants and giants.  First, from a strictly mathematical point of
view, the improved SNR, would allow even closer systems to be probed,
$a_{\min,0} = 0.2\,\au\times 10^{(H-18.5)/5}$.  However, the
combination of the growing size of the source and the shrinking
allowed orbit will quickly lead to ellipsoidal variations that are far
easier to detect compared to astrometric signals.  In such cases, the
astrometric orbits (at known periods) will confirm and refine the BH
detection.  On the other hand, for such luminous sources, all main-sequence
companions (i.e., up to $1\,M_\odot$) will be effectively dark.  This
leads to a threshold (for such equal-mass but very-unequal-luminosity
systems) of $a_\min = 0.4\,\au\times 10^{(H-18.5)/5}$
and $P_\min=65\,{\rm d}\times 10^{0.3(H-18.5)}$.  For example,
for an $H=16$ giant, the values would be $a_\min = 26\,R_\odot$ and
$P_\min =12\,$d.  The first number indicates that this system would already
be in or near the ellipsoidal regime (so vetted photometrically), but
the second shows that a large fraction of the short-period
BH-giant binary regime
would face serious contamination from main-sequence stars.  Nor would
there be much added sensitivity at the very shortest periods.
That is, nominally $a_{\min,0} = 13\,R_\odot$, but obviously a BH at this
separation would rip a giant apart.

I again emphasize that while there will overall be vast
``contamination'' for periods $0.25\,{\rm yr}\la P \la 5\,{\rm yr}$ in
the sense that the overwhelming majority of the successful orbital
fits will be due to stellar binaries, few of these are likely to
be confused with BHs.

\section{{BHs in $T=5\,{\rm yr}\la P\la 2T=10\,$yr Orbits}
  \label{sec:10yr}}

Under either the ``Gould'' or ``Roman'' observation strategies, BHs in
$P=2T=10\,$yr orbits will, in many cases be either
unambiguously recognized as BHs or be recognized as very strong
candidates, although few (if any) of them will have well-specified
orbits based on {\it Roman} astrometry alone.  Nevertheless, because
such recognition will immediately prompt RV observations, the
combination of astrometric data and RV followup will lead to
approximate solutions within a few years and precise solutions within
$\sim 5\,$yr from the completion of the {\it Roman} mission.  For
example, assuming a circular, face-on, $P=9.5\,$yr, orbit (so,
$a=9.7\,\au(M/10\,M_\odot)^{1/3}$), the star orbiting the BH
would deviate by $9.5\,\au$ or $\Delta\theta = 1.2\,\mas$
during the two middle seasons relative to what would be predicted
based on proper-motion alone as derived from the two wing seasons.
Such a large deviation could not be generated by any stellar binary in
the bulge.  The same observations would yield a precise source
parallax, so there would be no question of the system being in the
disk.

On the other hand for an edge-on orbit, such a maximal deviation would
only occur for the special case that the first epoch were at conjunction
or opposition, while there would be no deviation at all if the
first epoch were at quadrature.  At other inclinations, the situation
would be intermediate.  Hence, overall, some observed deviations would
be so large as to be inconsistent with stellar-binary contaminants, others
would lie deep in the ``astrophysical noise'' due to stellar binaries, and
a few would be on the boundary.

The situation for BH binaries orbiting in periods between $T\la P\la 2T$
would be overall similar, but for periods toward the shorter end of this
range, there would be increasingly precise orbital solutions.

At periods longer than $P>2T$ (and still for circular orbits),
the maximal deviation would drop as
$\sim 1.2\,\mas\,(M/10 M_\odot)^{1/3}(P/9.7\,{\rm yr})^{2/3}
(1-\cos((\pi/2)(9.7\,{\rm yr}/P))$,
which tends toward
$\rightarrow 1.5\,\mas\,(M/10 M_\odot)^{1/3}(P/9.7\,{\rm yr})^{-4/3}$
at large $P$.  This signal quickly merges into the background of
stellar binaries.  See Section~\ref{sec:expected-signal}.

Of course, not all BH-binaries $T\la P\la 2T$ can be detected in this
manner.  Highly eccentric systems will spend a long time at apocenter,
and these will give rise to much smaller deviations, which then can be
confused with stellar binaries.  Nevertheless, many can be found, and
most of the rest can be located by aggressive RV followup.
See Section~\ref{sec:expected-signal}.

\section{{Precision of Proper Motion, Parallax, Acceleration, and Jerk}
  \label{sec:precision}}

Changing the observational strategy according to
Equation~(\ref{eqn:roman-vs-ag}) will impact the survey's astrometric
and photometric sensitivity to many phenomena.  Fortunately, it will
have extremely minimal impact on the survey's central goal of making a planet
and free floating planet (FFP) census because, from the standpoint of
planetary anomalies, the different seasons can be considered independently.

However, it will affect the measurement precision of proper motions, 
which provide important auxiliary input into the interpretation of planetary
events.  Indeed, this is the reason for the adoption of the {\it Roman}
survey strategy given in Equation~(\ref{eqn:roman-vs-ag}), i.e., to
optimize these proper-motion measurements.

In this section, I analyze the impact of survey strategy on
the astrometric precision of proper motion ($\mu$), angular acceleration
($\alpha$), and angular jerk ($j$).  I report all three
results, and I specifically compare the performance of the {\it Roman}
and Gould strategies, as described by Equation~(\ref{eqn:roman-vs-ag}).
I give particular attention to the issue of proper motions because this
is the only one that impacts mission-critical performance.
I begin with the mathematical formalism, which follows \citet{gould03}.

\subsection{{Mathematical Formalism}
  \label{sec:math}}

For a linear function (of, say, time) of $p$ trial functions $f_i(t)$,
\begin{equation}
  F(t;a_1,\ldots,a_p) = \sum_{i=1}^p a_i f_i(t),
\label{eqn:foft}
\end{equation}
that is fitted to $N$ data points, the inverse covariance matrix of the $a_i$
is given by \citep{gould03}
\begin{equation}
  b_{ij} = \sum_{k,l=1}^N  f_i(t_k)B_{kl}f_j(t_l),
\label{eqn:bij}
\end{equation}
where $B_{kl}$ is the inverse covariance matrix of the data points.
Specializing to the case that the errors are all equal to $\sigma_0$
and also uncorrelated, i.e., $B_{kl} = \delta_{ij}/\sigma_0^2$,
Equation~(\ref{eqn:bij}) reduces to the more familiar
\begin{equation}
  b_{ij} = \sum_{k=1}^N   {f_i(t_k)f_j(t_k)\over \sigma_0^2}
\label{eqn:bij2}
\end{equation}

In most (though not all) of what follows, $F(t)$ will be a truncated $q$-th
order Taylor series (so, $p=q+1$)
\begin{equation}
  f_i(t) = {t^i\over i!}\qquad (i=0,\ldots q),
\label{eqn:foft2}
\end{equation}
in which case
\begin{equation}
  b_{ij} = {N\over \sigma_0^2}{\langle t^{i+j}\rangle\over i! j!};
  \qquad \langle t^r\rangle\equiv {1\over N}\sum_k t_k^r.
\label{eqn:bij3}
\end{equation}

One can always choose the zero-point of time such that $\langle t^1\rangle=0$.
However, if the observations are even in time around this zero point (as
will be the case in all of the applications below), then this expectation
vanishes for all odd $q$, i.e.,
\begin{equation}
  b_{ij} = {N\over \sigma_0^2}{\langle t^{i+j}\rangle\over i! j!}\quad
  [{\rm mod}(i+j,2)=0];
  \qquad b_{ij} = 0\quad [{\rm mod}(i+j,2)=1]
\label{eqn:bij4}
\end{equation}
I will mainly consider two cases, $q=1$ ($p=2$) and $q=3$ ($p=4$).
Both can be understood from the explicit evaluation:
\begin{equation}
  b_{ij} = {N\over\sigma_0^2}
  \left(\matrix{1 & 0 &\langle t^2\rangle/2 & 0\cr
    0 & \langle t^2\rangle & 0 &\langle t^4\rangle/6\cr
    \langle t^2\rangle/2 & 0 &\langle t^4\rangle/4 & 0\cr
    0 & \langle t^4\rangle/6 & 0 &\langle t^6\rangle/36}
  \right).
    \label{eqn:bijmat}
\end{equation}
Hence, for $q=1$, $b_{ij}^{q=1}$ is the just the diagonal matrix
\begin{equation}
  b_{ij}^{q=1} = {N\over\sigma_0^2}
  \left(\matrix{1 & 0 \cr
    0 & \langle t^2\rangle}
  \right).
    \label{eqn:bijmat2}
\end{equation}
For $m=3$, the matrix is block-diagonal (after permuting rows and columns)
and so can be decomposed into even and odd matrices
\begin{equation}
  b_{ij}^{\rm even} = {N\over\sigma_0^2}
  \left(\matrix{1 & \langle t^2\rangle/2 \cr
    \langle t^2\rangle/2  & \langle t^4\rangle/4}
  \right)\quad (i,j=0,2);\qquad
  b_{ij}^{\rm odd} = {N\over\sigma_0^2}
  \left(\matrix{\langle t^2\rangle & \langle t^4\rangle/6  \cr
    \langle t^4\rangle/6  & \langle t^6\rangle/36}
  \right)\quad (i,j=1,3).
    \label{eqn:bijmat3}
\end{equation}

Because 
\begin{equation}
  \langle (\delta\tau)^r \rangle = {(\Delta t)^r\over (r+1)2^r}\qquad
  (r\ \rm even),
\label{eqn:deltatau}
\end{equation}
and 0 otherwise, while the $\tau_i$ span a $T\sim 5\,$yr mission,
$\langle\tau\rangle/\langle\delta\tau\rangle\sim (T/\Delta t)^r\sim 25^r$.
Hence, I will mostly ignore the spread of the $\delta\tau$
distribution, and just treat it as $n$ observations all at $\tau_i$.
The one exception is that this term cannot be ignored in the estimate
of the parallax error, which enters the present study only because
it is correlated with the proper-motion measurement.
That is, with this one exception, I will be making the substitution
$\langle t^r\rangle \rightarrow \langle \tau^r\rangle$.  I thereby find,
\begin{equation}
  (\langle (t/Q)^2\rangle \ \langle (t/Q)^4\rangle \ \langle (t/Q)^6\rangle)=
  {1\over 3}\left(\matrix{155 & 9587 & 664715\cr 131 & 8963 & 649091}\right)
  \qquad {(Roman)\atop (\rm Gould)}.
  \label{eqn:roman-vs-ag2}
\end{equation}

\subsection{{Proper Motions}
  \label{sec:pms}}

\subsubsection{{\rm ppm}
  \label{sec:ppm}}

If we ignore parallax, as well as all higher order effects, such as
acceleration and jerk, then the proper motion (in each direction)
is derived from a 2-parameter fit to position and proper motion, i.e., ppm.
If the zero-point of time is set to the mission midpoint, so that
$\langle t\rangle \equiv \sum_{i,j} t_{i,j}/N = 0$, then \citep{gould03},
\begin{equation}
  \sigma(\mu) = {\sigma_0\over\sqrt{N}}\langle t^2\rangle^{-1/2}
  \rightarrow {\sigma_0\over\sqrt{N}}(\langle \tau^2\rangle^{-1/2}
  = {\sigma_0\over\sqrt{N}}
  \Biggl(
        {\sqrt{48/155}\atop \sqrt{48/131}}\Biggr){\rm yr}^{-1}
        \qquad {(Roman)\atop ({\rm Gould})},
\label{eqn:pm}
\end{equation}
where I have specifically evaluated this expression for the time series
that are described by Equations~(\ref{eqn:tij}) and (\ref{eqn:roman-vs-ag}).
Note that in the second step, I have simplified
$\langle t^2\rangle =\langle \tau^2 \rangle+\langle (\delta\tau)^2\rangle
\rightarrow \langle \tau^2\rangle$.  Had I not done so, the final expression
would have been higher by $155\rightarrow 155.16$ and $131\rightarrow 131.16$,
a difference of $10^{-3}$.
Thus, the observational schedule that I have proposed has a bigger
proper-motion error than the {\it Roman} schedule by a factor
$\sqrt{155/131} = 1.088$.  And this is presumably the reason that the
{\it Roman} schedule was adopted.  That is, even though the detailed
calculation presented here was probably not done, it is obvious that
by weighting the observations toward the wings of the mission, the proper-motion
precision will be optimized.  

\subsubsection{{\rm pppm}
  \label{sec:pppm}}

It is often reasonable to estimate the proper-motion precision by ignoring
parallax because the error made in doing so is typically small.  However,
in the present case, the difference in precisions between the {\it Roman}
and Gould schedules is also small, so parallax cannot be ignored.
Thus, the proper approach is to do a full 5-parameter pppm fit, wherein
the parallax intrinsically links the two components of the position and
proper motion.  The inverse covariance matrix is then given by,
\begin{equation}
  b_{ij} = {N\over \sigma_0^2} 
  \left(\matrix{ 1 & 0 & \langle t\rangle & 0 &\langle f_\pi\rangle \sin\beta\cr
  0 & 1 & 0 &\langle t\rangle  &\langle f_\pi\rangle \cos\beta\cr
  \langle t\rangle &  0 & \langle t^2\rangle & 0 & \langle f_\pi t\rangle \sin\beta\cr
  0 & \langle t\rangle & 0 & \langle t^2\rangle  & \langle f_\pi t\rangle \cos\beta\cr
  \langle f_\pi\rangle \sin\beta & \langle f_\pi\rangle \cos\beta & \langle f_\pi t\rangle \sin\beta & \langle f_\pi t\rangle \cos\beta & \langle f_\pi^2\rangle,
  }\right)
  \label{eqn:5parmfull}
\end{equation}
where the rows and columns are ordered in $(\lambda,\beta)$ ecliptic
coordinates $(\theta_\beta,\theta_\lambda,\mu_\beta,\mu_\lambda,\pi)$,
and
\begin{equation}
  f_\pi (t) = \cos\omega_\oplus t, \qquad \omega_\oplus\equiv {2\pi\over \rm yr},
  \label{eqn:fofpi}
\end{equation}
i.e., I ignore Earth's eccentricity.  One can always choose the time
zero-point so that $\langle t\rangle = 0$, and I have already done so in
Equations~(\ref{eqn:tij}) and (\ref{eqn:roman-vs-ag}).  Once this is done,
it does not automatically follow the $\langle f_\pi\rangle$ will also vanish.
However, for the case of either schedule, they do.  Thus, the first two rows
(and columns)
are (1 0 0 0 0) and (0 1 0 0 0), and they therefore decouple from the
rest of the matrix.  Next, I make the very small (0.1\%) approximation
that $\langle t^2\rangle = \langle \tau^2\rangle$, and I evaluate
\begin{equation}
\langle f_\pi t\rangle = \langle f_{\pi,0}(\tau)\tau\rangle \langle \cos(\omega_\oplus\delta\tau)\rangle, \qquad
\langle f_\pi^2\rangle = \langle \cos^2(\omega_\oplus\delta\tau)\rangle,\qquad
\end{equation}
where $f_{\pi,0}$ is $f_\pi(\tau_i)$ evaluated at the $m=6$ equinoxes, i.e.,
\begin{equation}
  f_{\pi,0} = (-1,+1,-1,+1,-1,+1),
  \label{eqn:fpi0}
  \end{equation}
for both the {\it Roman} and Gould
schedules. This yields the $3\times 3$
$(\mu_\beta,\mu_\lambda,\pi)$ matrix
\begin{equation}
  b_{ij}= {N\over\sigma_0^2}
  \left(\matrix{\langle \tau^2\rangle  & 0 & \langle f_{\pi,0}(\tau) \tau\rangle \langle\cos(\omega_\oplus\delta\tau)\rangle
    \sin\beta\cr
   0 & \langle \tau^2\rangle  & \langle f_{\pi,0}(\tau) \tau\rangle \langle\cos(\omega_\oplus\delta\tau)\rangle \cos\beta\cr
   \langle f_{\pi,0}(\tau) \tau\rangle \langle\cos(\omega_\oplus\delta\tau)\rangle \sin\beta & \langle f_{\pi,0}(\tau) \tau\rangle \langle\cos(\omega_\oplus\delta\tau)\rangle \cos\beta & \langle \cos^2(\omega_\oplus\delta\tau)\rangle
  }\right)
    \label{eqn:3parmfull}
\end{equation}

It is not difficult to proceed with Equation~(\ref{eqn:3parmfull}), and
I do so below.  Nevertheless, the basic physics of this equation become
much clearer if I make two simplifications, which I do temporarily.
First, I assume that the source lies exactly on the ecliptic, so that $\beta=0$.
This is actually nearly so for {\it Roman} sources.  Second, I assume
that $\cos(\omega_\oplus\delta\tau)\rightarrow 1$.  This would be true if the
{\it Roman} seasons were much shorter than they actually are.  Then
Equation~(\ref{eqn:3parmfull}) reduces to
\begin{equation}
  b_{ij}= {N\over\sigma_0^2}
  \left(\matrix{\langle \tau^2\rangle  & 0 & 0 \cr
    0 & \langle \tau^2\rangle  & \langle f_{\pi,0}(\tau) \tau\rangle \cr
   0 & \langle f_{\pi,0}(\tau) \tau\rangle  & 1}\right)
    \label{eqn:3parmreduced}
\end{equation}

The first row/column in Equation~(\ref{eqn:3parmreduced})
($\mu_\beta$) decouples from the other two.  Its solution is identical
to what was derived in Section~\ref{sec:ppm}.  The remaining matrix is
easily inverted to give the error in the other direction
\begin{equation}
  [\sigma(\mu_\lambda)]^2 = c_{22} = {\sigma_0^2\over N}\,
  {1\over \langle\tau^2\rangle - \langle f_{\pi,0}(\tau)\tau\rangle^2}.
    \label{eqn:c22eval}
\end{equation}
That is,
$c_{11}/c_{22} = 1 - \langle f_{\pi,0}(\tau)\tau\rangle^2/\langle \tau^2\rangle$.
One can easily evaluate $\langle \tau^2\rangle$ and
$\langle f_{\pi,0}(\tau)\tau\rangle$ using Equations~(\ref{eqn:roman-vs-ag})
and (\ref{eqn:fpi0}):
$(\langle\tau^2\rangle,\langle f_{\pi,0}\tau \rangle^2)_{Roman}=(155/3,(7/3)^2)Q^2$
and
$(\langle\tau^2\rangle,\langle f_{\pi,0}\tau \rangle^2)_{\rm Gould}=(131/3,1)Q^2$.
The reason for the much larger $\langle f_{\pi,0}\tau \rangle^2$ for {\it Roman}
is mathematically clear when one mentally calculates it.  For the
Gould schedule, each successive pair of equinoxes (when combined using the
$\pm 1$ cofactor, $f_{\pi,0}$) contributes $2Q$.  However, for {\it Roman}, the
second pair contributes $10Q$.  Physically, the long interval between
these two observed equinoxes induces correlations between the parallax and the
proper motion.  The {\it Roman} precision remains better than the Gould
precision, but for the $\lambda$ direction, its advantage is reduced from
a factor $\sqrt{155/131} = 1.088$ to $\sqrt{(155-16.33)/(131-3)} = 1.041$.
Averaged over all directions, the rms advantage is 1.065.

In fact, despite the seemingly crude approximations that I made,
the original matrix (Equation~(\ref{eqn:3parmfull})) yields almost
exactly the same result.  First note that
\begin{equation}
  \langle\cos(\omega_\oplus\delta\tau)\rangle =
             \sinc(\omega_\oplus \Delta t/2),\qquad
             \langle \cos^2(\omega_\oplus\delta\tau)\rangle =
                        {1+\sinc(\omega_\oplus \Delta t)\over 2}.
  \label{eqn:sincs}
\end{equation}
The determinant of Equation~(\ref{eqn:3parmfull}) can then
be expressed as
\begin{equation}
  {\rm det}(b) = {N\over \sigma_0^2}
  \langle \tau^2\rangle^2 \langle \cos^2(\omega_\oplus\delta\tau)\rangle
  (1 - \epsilon\zeta(\omega_\oplus\Delta t/2)),\qquad \epsilon\equiv
          {\langle f_{\pi,0}(\tau)\tau\rangle\over
            \langle \tau^2\rangle}
    \label{eqn:3parmfulldet}
\end{equation}
where
\begin{equation}
  \zeta(\psi) =  {2\,\sinc^2(\psi)\over \sinc (2\psi)};
  \qquad \zeta(\omega_\oplus \Delta t/2) \rightarrow 0.9965.
    \label{eqn:zetadef}
\end{equation}
Hence, the errors in the two components of the proper motion are
given by
\begin{equation}
  {[\sigma(\mu_\beta)]^2\atop [\sigma(\mu_\lambda)]^2}=
  {c_{11}\atop c_{22}} = \Biggl[{\sigma_0^2/N\over
  \langle\tau^2\rangle(1-\epsilon\zeta)}\Biggr]
  {(1-\epsilon\zeta \sin^2\beta)\atop (1-\epsilon\zeta \cos^2\beta)}
  \rightarrow \Biggl[{\sigma_0^2/N\over
      \langle\tau^2\rangle(1-\epsilon)}\Biggr]
    {(1-\epsilon\sin^2\beta)\atop (1-\epsilon\cos^2\beta)},
    \label{eqn:3parminv}
\end{equation}
and thus the rms error over all directions is
\begin{equation}
\overline{\sigma(\mu)} = \sqrt{c_{11}+c_{22}\over 2}
  = {\sigma_0\over \sqrt{N}}\sqrt{1-\epsilon/2\over
  \langle\tau^2\rangle(1-\epsilon)}\rightarrow
   {\sigma_0\over \sqrt{N\langle\tau^2\rangle}}\biggl(1+{\epsilon\over 4}\biggr).
    \label{eqn:3parminvave}
\end{equation}

Using this formula, I find that the average proper motion precision for the
Gould schedule is 6.6\% worse than for the {\it Roman} schedule, essentially
identical to the result obtained using the crude approximations of the
previous section.

\subsection{{Precision of Acceleration Measurement}
  \label{sec:acc}}

Next, I evaluate the error in the astrometric acceleration, $\alpha$ .
I invert Equation~(\ref{eqn:bijmat3}) and substitute in
Equation~(\ref{eqn:roman-vs-ag2}) to obtain
\begin{equation}
  \sigma(\alpha) = \sqrt{c_{22}} = \sqrt{b_{00}\over b_{00} b_{22}-b_{02}^2}
  ={\sigma_0\over\sqrt{N}}\Biggl(
  {\sqrt{72/37}\atop \sqrt{72/76}}\Biggr){\rm yr}^{-2}
  \qquad {(Roman)\atop (\rm Gould)}.
\label{eqn:acc}
\end{equation}
Hence, contrary to the case of proper motion, for acceleration, it is
the {\it Roman} schedule that induces larger errors, and here by a larger
factor: $\sqrt{76/37} = 1.43$.

\subsection{{Precision of Jerk Measurement}
  \label{sec:jerk}}

The jerk, $j$, is the first higher-order term after the acceleration.
I invert Equation~(\ref{eqn:bijmat3}) and substitute in
Equation~(\ref{eqn:roman-vs-ag2}) to obtain
\begin{equation}
  \sigma(j) = \sqrt{c_{33}} = \sqrt{b_{11}\over b_{11} b_{33}-b_{31}^2}
  ={\sigma_0\over\sqrt{N}}\Biggl(
  {(8/7)\sqrt{930/197}\atop 4\sqrt{786/1019}}\Biggr){\rm yr}^{-3}
  ={\sigma_0\over\sqrt{N}}\Biggl({2.48
  \atop 3.51}\Biggr){\rm yr}^{-3}
  \qquad {(Roman)\atop (\rm Gould)}.
\label{eqn:jerk}
\end{equation}

Hence, for jerk, almost exactly contrary to acceleration, the original
{\it Roman} schedule leads to 1.41 times better precision than my proposed
revision.

\subsection{{Evaluations for {\it Roman} Experiment}
  \label{sec:eval}}

Adopting the astrometric precision from Figure~1 of \citet{gould15}
($\sigma_0=0.5\,\mas\times 10^{(H_{\rm Vega}-18.5)/5}$) and $N=4\times 10^4$
epochs, and applying it to the above results according my proposed schedule, I
obtain
\begin{equation}
  \sigma(\mu) = 1.4{\mu{\rm as}\over \rm yr}\times 10^{(H-18.5)/5},\quad
  \sigma(\alpha) = 2.4{\mu{\rm as}\over \rm yr^2}\times 10^{(H-18.5)/5},\quad
  \sigma(j) = 8.8{\mu{\rm as}\over \rm yr^3}\times 10^{(H-18.5)/5}.
\label{eqn:eval}
\end{equation}

\section{{Expected BH Acceleration and Jerk Signals}
  \label{sec:expected-signal}}

To illustrate the scale of the expected signal, I consider face-on
circular orbits, and I approximate the companion as being a test
particle.  I also assume that the system is at $D_s = 8\,\kpc$.
Then for long orbits $P\gg T$, the system will effectively remain at a fixed
phase $\phi$ during the observations, so that the ``instantaneous''
acceleration and jerk will be:
\begin{equation}
  \alpha =
  123{\mu{\rm as}\over \rm yr^2}\biggl({M \over 10\,M_\odot }\biggr)^{}
  \biggl({a \over 20\,\au }\biggr)^{-2};\qquad
  27{\mu{\rm as}\over \rm yr^3}\biggl({M \over 10\,M_\odot }\biggr)^{3/2}
  \biggl({a \over 20\,\au}\biggr)^{-7/2},
\label{eqn:eval2}
\end{equation}
in the radial and tangential directions, respectively.

Comparing Equation~(\ref{eqn:eval2}) with Equation~(\ref{eqn:eval}),
it is clear that BHs in orbits out to $a\sim 50\,\au$ will be detectable
with SNR of at least $10\,\sigma$ in acceleration
and out to $a\sim 15\,\au$ in jerk.

\section{{Astrophysical Background}
  \label{sec:background}}

However, the majority of the BH systems ``detected'' via such acceleration
and/or jerk measurements would be swamped by similar signals from ordinary
binaries.

For example, consider a BH system in a face-on orbit with
$M=10\,M_\odot$, $a=30\,\au$.  It would have $P\sim 50\,$yr,
$\alpha=55\,\mu{\rm as\,yr}^{-2}$ and $j=6.6\,\mu{\rm as\,yr}^{-3}$.
That is, a strongly-detected acceleration and a completely undetectable jerk.
The acceleration could be mimicked by an
$M=0.7\,M_\odot$, $a=8\,\au$ WD binary, which have a period, $P=17\,$yr.  If
this putative system were modeled as also being observed face-on, then
it would have a jerk $j=20\,\mu{\rm as\,yr}^{-3}$, and the absence of such a
2.3-$\sigma$ ``signal'' might be regarded as evidence against this WD
interpretation.  However, if the same system were modeled as being edge-on,
with the instantaneous jerk vector pointed at the observer, then
this would also predict vanishing astrometric jerk.  As there are orders of
magnitude fewer BHs than WDs (not to speak of $0.7\,M_\odot$ K dwarfs, which
would yield a similar signal), the acceleration and jerk ``signals''
of this BH would not be regarded as particularly promising.

Nevertheless, there will be a transition region of BH-binaries
between such acceleration-only detections and the $P\la T$ full orbits
that are illustrated in Figure~\ref{fig:gaps} that will appear promising.
If, for example, simulations show that they have a 10\% probability of
being BH-binaries, then this could be verified by RV followup observations.
Assuming that such 10\% estimates were correct, one would have to
monitor 10 candidates to find one BH binary.

\begin{figure}
\plotone{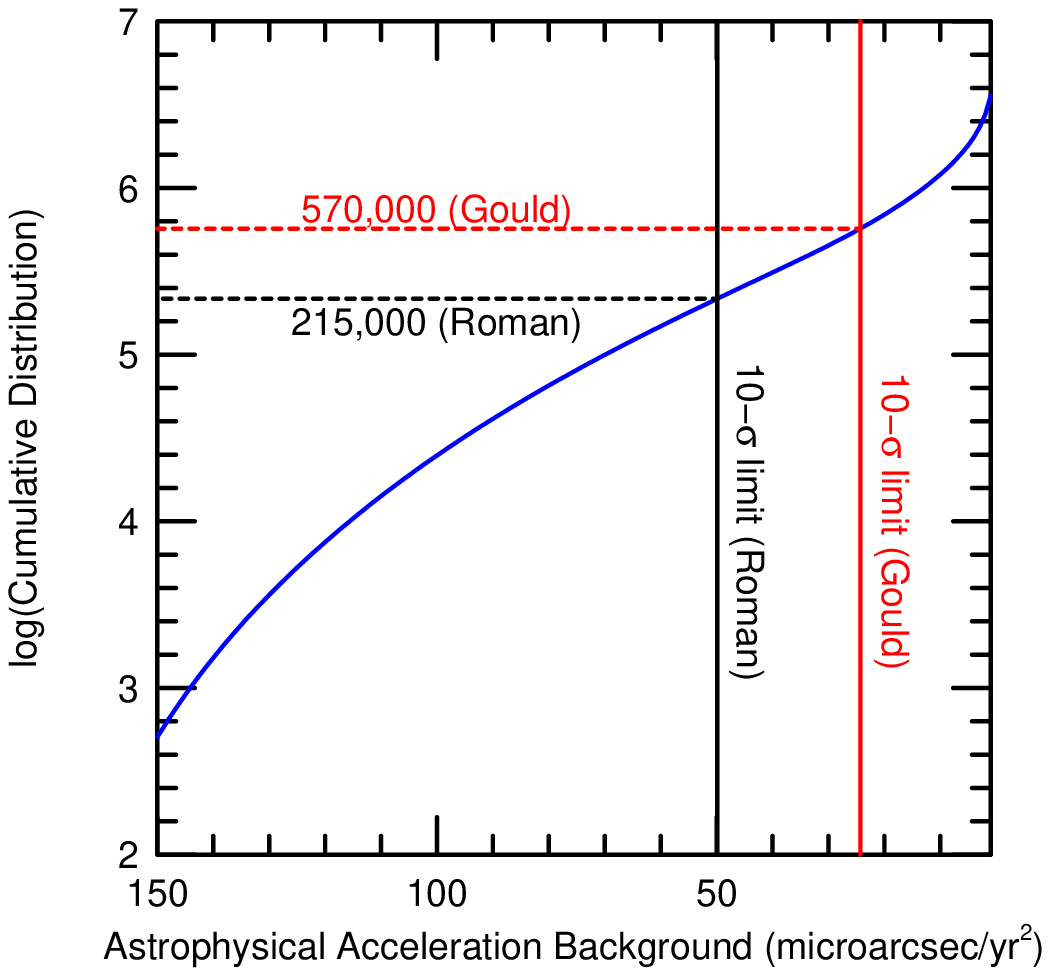}
\caption{Astrophysical acceleration background due to stellar binaries,
  assuming 30 million total sources, $H_{\rm Vega}<18.5$, and G-dwarf
  binary statistics derived from \citet{dm91}.  The cumulative
  distribution is shown in blue, while the 10-$\sigma$ detection thresholds
  are shown in red and black for the current (``Roman'') and proposed
  (``Gould'') observing strategies, respectively.  The Gould threshold
  reaches to BH binaries with periods $P<95\,{\rm yr}(M/10 M_\odot)^{1/4}$,
  but for periods $P\ga 2 T=10\,$yr, one must comb through 570,000
  astrophysical background signals with RV monitoring.  As discussed
  in Section~\ref{sec:background}, this could be carried out with
  a 1000-fiber multi-object spectrograph, beginning either immediately
  after the {\it Roman} mission, or perhaps with some delay.
  }
\label{fig:back}
\end{figure}

Moreover, one could imagine a much more aggressive followup program
based on, say, a 1000-fiber multi-object spectrograph.  To provide
a concrete understanding of how such a program would work, I
show in Figure~\ref{fig:back} the cumulative astrophysical background noise
due to stellar binaries as a function of observed acceleration.  I derive
these estimates from the periods and companion masses of G dwarfs as
presented by \citet{dm91} and assume circular orbits viewed at random
orientations.  I restrict the sample to periods $P>10\,$yr.
The vertical bars represent the 10-$\sigma$ detection
limits for the Gould (red) and {\it Roman} (black) observing schedules.
The plot is scaled to 30 million total sources.  As indicated, there
are 570,000 and 215,000 stellar-background sources that meet the Gould
and {\it Roman} thresholds respectively.  The Gould 10-$\sigma$ limit
corresponds to periods $P=95\,{\rm yr} (M/M_\odot)^{1/4}$.

Using a 1000-fiber system, one could monitor all 570,000 candidates
with $H\la 18.5$ sources and $\geq 10$-$\sigma$ acceleration signals
with just 570 pointings.  Because periods $P\la T$
would already be ruled out, such monitoring could take place at a
``Nyquist'' cadence of $\Gamma = 2/T= 0.4\,{\rm yr}^{-1}$, so that
only 230 pointings would be required per year.  After 3 or 4 epochs,
most candidates would be either ruled out as due to stellar binaries
(so removed from the list of candidates) or constrained to be of longer
periods (leading to a reduction of the ``Nyquist'' cadence).  Hence,
the number of pointings per year could be greatly reduced.

While such an RV-followup program would represent a considerable
expenditure of resources, these would be tiny compared to an ab initio
RV survey, which would be the only other way to find bulge BH-binary
systems in wide orbits.

\section{{Other Considerations}
  \label{sec:other}}

While my advocacy for the Equation~(\ref{eqn:roman-vs-ag}) scheduling change 
was motivated only by goal of improving sensitivity to BH binaries
in $P\la T=5\,$yr orbits, there are several other reasons to make
such a change.

First, the sensitivity to wide-orbit ``acceleration-only'' candidates
that was just discussed in Sections~\ref{sec:expected-signal} and
\ref{sec:background}, will be
increased by a factor $\sqrt{76/37}=1.43$ (see Section~\ref{sec:acc}),
which would push to larger semi-major axes (at fixed mass) by a factor
1.2, and so to larger periods by a factor 1.7.  While this ``greater
sensitivity'' would lead to few if any immediate BH-binary detections,
it would lay the basis for a future RV-followup survey that could
comprehensively study such wide-orbit systems.

Second, my proposed schedule would be more robust against catastrophic
mission failure.  The experience of the {\it Kepler} mission should
remind us that such catastrophic failures do happen.  While these can,
in principle, happen at any time, their probability grows with time
simply because critical systems can prematurely wear out.  For example,
according to the current plan, filter changes by the microlensing program
alone will exhaust half the planned endurance of the filter wheel.
Thus, while it is possible that a catastrophic failure would occur after
2 years (in which case the {\it Roman}/Gould schedules would carry out
(3/2) seasonal campaigns), it is more likely that the failure would
occur after 2.5 or 3 years, in which case the corresponding comparisons
would be (3/3) and (3/4) seasonal campaigns.  For any of these
possibilities, the proper-motion precision would be seriously degraded
for either the {\it Roman} of Gould schedule.  However, considering
only the ``fairest'' comparison, in which both schedules observe for
3 seasons [$(-9,-7,-5)$ versus $(-9,-7,-1)$], the Gould schedule
has better proper motion precision by a factor $\sqrt{13/3}\sim 2.1$,
which would be very significant.

Third, I conjecture that the Gould schedule would be overall more sensitive
to astrometric microlensing signals from BHs.


.\section{{Summary}
\label{sec:summary}}

There are two central points to this paper.  First, assuming that
{\it Roman} astrometry can achieve photon-limited precision, the {\it Roman}
microlensing survey will return BH mass measurements for BHs in binaries
with solar-mass stars (of which there are about 30 million) with periods
$10\,{\rm d}\la P \la 5\,$yr.  These measurements can be verified and refined
using RV followup.

Second, the current {\it Roman} scheduling (3 seasons at the beginning of
the mission and 3 at the end) is sub-optimal for this purpose.  Hence,
I advocate a different allocation of the 6 seasons,
i.e., 2 ON, 2 OFF, 2 ON, 2 OFF, 2 ON.

I also showed that this change would degrade the proper-motion measurement
precision (the main driver of the current schedule) by only 4\%-9\%, depending
on the orientation of the proper-motion vector relative to the ecliptic.

And, in addition, I showed that while BHs in many-decade orbits could not
be reliably identified in the {\it Roman} survey itself, {\it Roman} could
identify candidate wide BH-binary systems that could be securely selected
in a future massive multi-fiber RV campaign.

\acknowledgments
I thank Subo Dong for valuable discussions.

\clearpage

\begin{thebibliography}{99}

\bibitem[Cassan et al.(2022)]{cassan22} Cassan, A., Ranc, C., Absil, O., et al., 2022, Nature Astronomy, 6, 121


\bibitem[Delplancke et al.(2001)]{delplancke01}Delplancke, F., G\'orski, K.M., \& Richichi, A. 2001, \aap, 375, 701
  
\bibitem[Dong et al.(2019)]{kojima1} Dong, S., M\'erand, A., Delplancke-Strobale, F. 2019, \apj, 871, 70

\bibitem[Duquennoy \& Mayor(1991)]{dm91}Duquennoy, A., \& Mayor, M. 1991, \aap, 248, 485
  
\bibitem[El-Badry et al.(2023a)]{elbadry23a}El-Badry, K., Rix, H.W., Quataert, I., et al.\ 2023a, \mnras, 518, 1057
  
\bibitem[El-Badry et al.(2023b)]{elbadry23b}El-Badry, K., Rix, H.W., Cendes, Y., et al.\ 2023b, \mnras, 521, 4323
  
%
  
\bibitem[Gaia Collaboration et al.(2024)]{mazeh24}Gaia Collaboration, Panuzzo, P., Mazeh, T., et al.\ 2024, \aap, 686L, 2

%
  
\bibitem[Gould(2003)]{gould03} Gould, A.\ 2003, 2003astro.ph.10577 
  
\bibitem[Gould(2023)]{industrial} Gould, A.\ 2023, arXiv:2310.19164   
  
%
  
\bibitem[Gould \& Yee(2014)]{gouldyee14} Gould, A. \& Yee, J.C.\ 2014, \apj, 784, 64

%

\bibitem[Gould et al.(2015)]{gould15} Gould, A., Huber, D., Penny, M., \& Stello D.\ 2015, JKAS, 48, 93
  
\bibitem[Gould et al.(2023)]{kb222397} Gould, Ryu, Y.-H., Yee, J.C., et. al., 2023, \aj, 166, 100
  
%
  
\bibitem[Hog et al.(1995)]{hnp95} Hog, E., Novikov, I.D., \& Polanarev, A.G. 1995, \aap, 294, 287

%
  
\bibitem[Lam et al.(2022)]{ob110462b} Lam, C.Y., Lu, J.R., Udalski, A., et al., 2022, \apjl, 933, L23

%

\bibitem[Miyamoto \& Yoshii(1995)]{my95}Miyamoto, M. \& Yoshii, Y. 1995, \aj, 110, 1427

%

\bibitem[Mr\'oz et al.(2022)]{ob110462c} Mr\'oz, P, Udalski, A., \& Gould, A., 2022, \apjl, 937, L24 

%

\bibitem[Sahu et al.(2022)]{ob110462a} Sahu, K.C., Anderson, J., Casertano, S., st al. 2022, \apj, 933, 83
  
%

\bibitem[Walker(1995)]{walker95} Walker, M.A. 1995, \apj, 453, 37


\end{thebibliography}
\end{document}